\documentclass[sn-mathphys,Numbered]{sn-jnl}

\usepackage{float}
\usepackage{graphicx}
\usepackage{multirow}
\usepackage{amsmath, amssymb, amsfonts}
\usepackage{amsthm}
\usepackage{mathrsfs}
\usepackage[title]{appendix}
\usepackage{xcolor}
\usepackage{textcomp}
\usepackage{manyfoot}
\usepackage{hyperref}
\usepackage{cleveref}
\usepackage{booktabs}
\usepackage{algorithm}
\usepackage{algorithmicx}
\usepackage{algpseudocode}
\usepackage{listings}
\usepackage{indentfirst}
\usepackage{parskip} 
\usepackage{caption}
\usepackage{subcaption}

\begin{document}

\title{Quantum mechanics of particles constrained to spiral curves with application to polyene chains}

\author[1]{\fnm{Eduardo V. S.} \sur{Anjos}}\email{eduardo.anjos@ufpe.br}
\author[1]{\fnm{Antonio C.} \sur{Pav\~{a}o}}\email{antonio.pavao@ufpe.br}
\equalcont{These authors contributed equally to this work.}
\author*[2]{\fnm{Luiz C. B.} \sur{da Silva}}\email{LBarbosaDaSilva001@dundee.ac.uk}
\equalcont{These authors contributed equally to this work.}
\author[3]{\fnm{Cristiano C.} \sur{Bastos}}\email{cristiano.bastos@ufrpe.br}
\equalcont{These authors contributed equally to this work.}

\affil[1]{\orgdiv{Department of Fundamental Chemistry}, \orgname{Federal University of Pernambuco}, \orgaddress{\city{Recife, 50740-540}, \country{Brazil}}}
\affil[2]{\orgdiv{Division of Mathematics, School of Science and Engineering}, \orgname{University of Dundee}, \city{Dundee DD1 4HN}, \country{UK}}
\affil[3]{\orgdiv{Department of Chemistry}, \orgname{Rural Federal University of Pernambuco}, \orgaddress{\city{Recife, 52171-900}, \country{Brazil}}}

\abstract{\textbf{Context:} Due to advances in synthesizing lower dimensional materials there is the challenge of finding the wave equation that effectively describes quantum particles moving on 1D and 2D domains. Jensen and Koppe and Da Costa independently  introduced a confining potential formalism showing that the effective constrained dynamics is subjected to a scalar geometry-induced potential; for the confinement to a curve, the potential depends on the curve's curvature function.

\textbf{Method:} To characterize the $\pi$ electrons in polyenes, we follow two approaches. First, we utilize a weakened Coulomb potential associated with a spiral curve. The solution to the Schr\"{o}dinger equation with Dirichlet boundary conditions yields Bessel functions, and the spectrum is obtained analytically. We employ the particle-in-a-box model in the second approach, incorporating effective mass corrections. The $\pi$-$\pi^*$ transitions of polyenes were calculated in good experimental agreement with both approaches, although with different wave functions.}


\maketitle

\section{Introduction}
\label{sec::Intro}

Thanks to tremendous advances in experimental techniques, synthesizing lower dimensional materials became a reality. (See Ref. \cite{Chem2D2017} and references therein.) Such materials often display formidable properties that offer countless opportunities. With such advances comes the challenge of finding the wave equation that effectively describes quantum particles moving on 1D and 2D materials. To find the effective wave equation for a particle confined to move on a lower-dimensional region, it is necessary to account for the uncertainty relations since any confinement involves the full knowledge of the degrees of freedom associated with the motion along the direction orthogonal to the constraining region. In the 1950s, De Witt attempted to describe quantum confinement in a curved space through a quantization procedure, which resulted in an ordering ambiguity \cite{dW57}. A formalism that does not suffer from this ambiguity has been proposed independently by Jensen and Koppe \cite{JensenKoppe1971} in the 1970s and by Da Costa in the 1980s \cite{DaCosta1981}: their formalism shows that the effective constrained dynamics is subjected to a scalar geometry-induced potential. Jensen and Koppe analyzed a case where confinement occurs between two parallel surfaces. They obtained that the Schrödinger equation depends on a geometry-induced potential $V_{gip}$ that incorporates the geometry of the confinement region \citep{JensenKoppe1971}. On the other hand, Da Costa arrived at the same result by employing an explicit strong confining potential to restrict the particle's motion to the desired lower-dimensional region \citep{DaCosta1981}; for the confinement of a quantum particle to a curve, he obtained a Hamiltonian whose geometry-induced potential depends on the curve's curvature function. 

Several works have exploited the Jensen-Koppe-Da Costa's formalism. For example, there are studies of charge transport in semiconductors or carbon nanostructures \citep{DelCampo2014, DaSilva2017, Lima2021}. Del Campo \emph{et al.} studied geometry-induced potentials that result in better transmittance in bent waveguides \citep{DelCampo2014}. Da Silva \emph{et al.} studied the problem of prescribed geometry-induced potential for invariant surfaces, showing that the probability density distribution can be controlled if we add an extra charge to the surface \citep{DaSilva2017}. Lima \emph{et al.} calculated the energy and analyzed the implications of the geometry-induced potential for confinement in a helix, catenary, helicoid, and catenoid, concluding that for the helix, the angular momentum is quantized due to the geometry and that in the other cases, a continuous energy band of excited states appears \citep{Lima2021}. Experimentally, Onoe \emph{et al.} reported the observation of effects due to the geometry-induced potential on the Tomanaga-Luttinger liquid exponent in a 1D metallic $C_{60}$ polymer with an uneven periodic peanut-shaped structure \cite{Onoe12GeomEffectsPeanutTube}. As an alternative to Jensen-Koppe-Da Costa's formalism, Bastos \emph{et al.} studied the effects of intrinsic geometry on a particle confined in a generalized cylinder with a smooth cross-section. They noted that the topology, and not the geometry of the cross-section, plays a fundamental role in solving the problem in a Möbius strip and aromatic molecules \citep{Bastos2016}.

Electrons moving in the ballistic regime are less affected by the lattice structure and can often be described as free particles if we properly renormalize their mass \cite{Oshikiri1996}, thus giving rise to the concept of effective mass. A similar situation happens for $\pi$ electrons, i.e., electrons on a $\pi$ bond. The $\pi$ bonds are usually weaker than sigma bonds. Consequently, $\pi$ electrons can sometimes be reasonably described as a particle in a box \cite{ruedenberg1953, scherr1953, platt1953}. The $\pi$ electron wave functions are the usual trigonometric functions in such a regime.

In this work, we provide an alternative description of $\pi$ electrons by modeling them confined to a spiral-like curved 1D box. By incorporating a spiral behaviour, the new wave functions are given by certain Bessel functions. Bessel functions exhibit a more complex behaviour than the usual trigonometric functions and introduce new factors, such as zero modes and wave amplitude dependence on the energy level. We apply this idea to characterize the $\pi$ electrons in linear polyenes chains (Fig. \ref{fig:polyenes}). Specifically, we solve the Schr\"{o}dinger equation for a particle confined in certain spiral curves that can describe the 1D hydrogen atom and polyene linear chains.  Our findings suggest a correlation between the electronic confinement, the effective mass, and the geometry-induced potential.

This work is divided as follows. Section \ref{sec::1Dconfinement} presents Jensen-Koppe-Da Costa's formalism for quantum particles constrained to move on a plane curve. Section \ref{sect::1dHatom} discusses the geometry of plane curves with power-law curvature functions and applies it to the 1D hydrogen atom seen as a constrained quantum dynamics problem. Section \ref{sect::PolyeneCurve} introduces the family of plane curves that will be used to model $\pi$ electrons on polyene chains and presents the corresponding energy spectrum. Finally, in Section \ref{sect::Conclusion}, we present our concluding remarks.


\section{Constrained quantum dynamics on plane curves}
\label{sec::1Dconfinement}

\begin{figure}[t]
    \centering
 \includegraphics[width=0.6\linewidth]{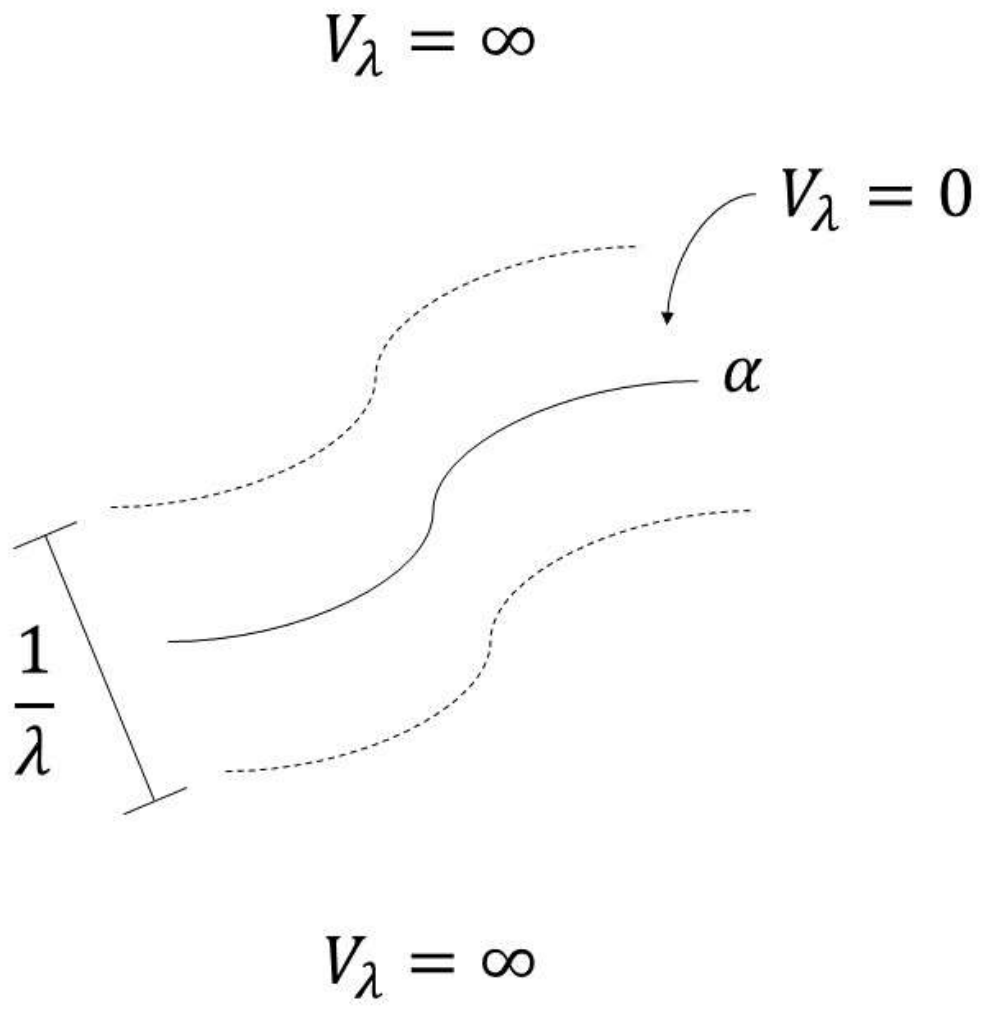}
    \caption{Particle-in-a-box constraining potential in the confinement of a quantum particle to a plane curve $\alpha:[a,b]\subseteq\mathbb{R}\to\mathbb{R}^2$. The particle is subject to a potential $V_{\lambda}$ that is zero on the points whose distance from $\alpha$ is smaller than or equal to $\frac{1}{2}\lambda^{-1}$, and infinity if otherwise. In the limit $\lambda\to\infty$, one obtains the behaviour \eqref{eq::ConstrainingPotential}.}
    \label{fig:PIBconfinement}
\end{figure}

Assume we want to describe the motion of a quantum particle of mass $m$ constrained to move along a plane curve $\alpha:[a,b]\to\mathbb{R}^2$. To find the equations for the constrained dynamics, we could follow Jensen and Koppe \cite{JensenKoppe1971} and describe the confinement by starting from the dynamics in the region between two neighbouring parallel curves and imposing homogeneous boundary conditions along them. If we denote the distance between the two neighbouring curves by $1/\lambda$, then taking the limit $\lambda\to\infty$, one obtains the equations that govern the constrained dynamics. In other words, Jensen and Koppe considered the confinement via a particle in a box model: the particle is subject to a potential $V_{\lambda}$ such that $V_{\lambda}(\vec{r})=0$ if the distance from $\vec{r}\in\mathbb{R}^2$ to $\alpha$ is smaller than or equal to $\frac{1}{2}\lambda^{-1}$, and $V_{\lambda}(\vec{r})=\infty$ if otherwise. (See Fig. \ref{fig:PIBconfinement}.)

Alternatively, following Da Costa \cite{DaCosta1981}, we may apply a family of explicit strong confining potentials $V_{\lambda}$ to restrict the particle's motion to the desired curve:
\begin{equation}\label{eq::ConstrainingPotential}
 \lim_{\lambda\to\infty} V_{\lambda}(\vec{r}) = \left\{
\begin{array}{ccc}
0 & , & \vec{r} \in \alpha\\ [4pt]
\infty & , & \vec{r} \not\in \alpha\\
\end{array}
\right..    
\end{equation}
These procedures allow us to decouple the tangential and normal degrees of freedom in the limit $\lambda\to\infty$. In other words, one separates the Hamiltonian into a term that governs the low energy motion in the tangent direction, which is the effective Hamiltonian along the constraint region, and a high energy motion in the normal direction. 

Employing the Jensen-Koppe-Da Costa formalism is necessary to account for the uncertainty relations since any confinement involves the full knowledge of some degrees of freedom, namely the motion along the directions orthogonal to the constraining region. De Witt's description of quantum confinement in a curved space resulted in an ordering ambiguity \cite{dW57}. The Jensen-Koppe-Da Costa formalism does not suffer from this ambiguity. In addition, it shows that the effective constrained dynamics is subjected to a scalar geometry-induced potential $V_{gip}$. 

The (effective) Schr\"{o}dinger equation for a quantum particle constrained to move on a plane curve $\alpha(s) = (x(s), y(s))$ is given by \cite{DaCosta1981}
\begin{equation}\label{eq::Hgip1D}
 -\frac{\hbar^2}{2m}\frac{d^2\psi(s)}{ds^2} +V_{gip}\,\psi(s) = E\psi(s), \quad V_{gip} = -\frac{\hbar^2}{8m}k(s)^2,
\end{equation}
where $s$ denotes the arc-length parameter, i.e., $\alpha'(s)\cdot\alpha'(s)=1$, and $k$ is the curvature function; $k(s)=\Vert\alpha''(s)\Vert$.

If we were to describe the constrained quantum dynamics on a plane curve $\alpha$ without the geometry-induced potential, then the resulting energy spectrum would not depend on the geometry of $\alpha$ but only on whether $\alpha$ is a closed or an open curve \cite{Bastos2012}. Indeed, solving Eq. \eqref{eq::Hgip1D} without $V_{gip}$ with homogeneous boundary conditions (open curve of length $L$) gives
\begin{equation}\label{PIBspectrumOpenCurve}
    E_{n}^{\textrm{op}} = \frac{h^2n^2}{8mL^2},
\end{equation}
while solving Eq. \eqref{eq::Hgip1D} without $V_{gip}$ with periodic boundary conditions (closed curve of length $L$) gives
\begin{equation}\label{PIBspectrumClosedCurve}
 E_{n}^{\textrm{cl}} = 4E_{n}^{\textrm{op}} = \frac{h^2n^2}{2mL^2}.
\end{equation}


\section{1D Hydrogen atom as a constrained quantum dynamics problem}
\label{sect::1dHatom}

If we consider a function of the form  $k(s) =\frac{1}{\sigma\sqrt{s}}$, where $\sigma$ is a real parameter, then confining a particle to move on a plane curve with curvature $k$ can lead to a geometry-induced corresponding to 1D Hydrogen atom:
\begin{equation}\label{Vgip1dHatom}
 V_{gip} = -\frac{\hbar^2}{8m}k(s)^2 = -\frac{\hbar^2}{8m\sigma^2s} = -K\frac{q_1q_2}{s},
\end{equation}
where $K$ denotes the permittivity of free space, $K=9\times10^9 Nm^2/C$, $q_1$ denotes the charge of the nucleus, and $q_2$ the charge of the electron. Indeed, take the constant $\sigma = \sqrt{\frac{\hbar}{8mKq_1q_2}}$. We shall refer to a curve $\alpha_H:[a,b]\to\mathbb{R}^2$ whose corresponding curvature function $k$ satisfies Eq. \eqref{Vgip1dHatom} as a \emph{Hydrogen curve}.

\subsection{Plane curves with power-law curvature function}

A plane curve $\alpha(s)$ with curvature function $k(s)$ can be parametrized as \citep{DoCarmo1976}
\begin{equation}
 s \mapsto \alpha(s)=\left(\int_{s_0}^s \cos(\int_{s_0}^v k(u) \, du) \, dv,\int_{s_0}^s \sin(\int_{s_0}^v k(u) \, du) \, dv\right).
\end{equation}
For a generic parametrization $\alpha(t)$, the arc length parameter can be obtained as a function of $t$ by the expression $s=\int_{t_0}^t\Vert\dot{\alpha}(\tau)\Vert\,d\tau$, while the curvature function is $k=\Vert\dot{\alpha}\times\ddot{\alpha}\Vert/\Vert\dot{\alpha}\Vert^3$.

The Hydrogen curve belongs to the family of curves with a power-law curvature function
\begin{equation}\label{eq::PowerLawCurvature}
    k(s) = \frac{1}{\sigma s^p}, \quad \sigma>0 \quad \mbox{and} \quad p\in\mathbb{R}.
\end{equation}
The Hydrogen curve $\alpha_H$ has $p=1/2$. Power-law curvature functions lead to spiral-like curves. 

Every plane curve satisfies the Frenet equations
\begin{equation}\label{eq::FranetFrameEqs}
\left\{\begin{array}{ccc}
\mathbf{t}'(s) & = & k(s)\,\mathbf{n}(s)\\
\mathbf{n}'(s) & = & -k(s)\,\mathbf{t}(s)\\
\end{array}
\right.\,,
\end{equation}
where $\mathbf{t}=\alpha'$ denotes the curve's unit tangent and $\mathbf{n}$ is the principal normal vector field. The geometric interpretation of the vector fields $\mathbf{t}$ and $\mathbf{n}$ is as follows. If we think of a plane curve as describing the motion of a particle in the plane, the unit normal $\mathbf{n}$ points in the direction of the centripetal acceleration vector. Indeed, applying the chain rule, $\frac{d\alpha}{dt}=v\frac{d\alpha}{dt}$, $v=\Vert d\alpha/dt\Vert$, from which we obtain that $\frac{d^2\alpha}{dt^2} = \frac{dv}{dt}\mathbf{t} + v^2k\,\mathbf{n}$. 

The solutions of Eq. (\ref{eq::FranetFrameEqs}) for the power-law case, Eq. \eqref{eq::PowerLawCurvature}, are given by 
\[
\mathbf{t}(s) = \mathbf{a}\, C_p(s) + \mathbf{b}\, S_p(s)\quad \mbox{and} \quad 
\mathbf{n}(s) = \sigma s^p\,\mathbf{t}'(s) = -\mathbf{a}\,S_p(s)+\mathbf{b}\,C_p(s),
\]
where $\mathbf{a}$ and $\mathbf{b}$ are constant vectors and the real functions $C_p$ and $S_p$ are defined as
\begin{equation}\label{eq::DefOfCpAndSp}
C_p(s) = \left\{
\begin{array}{ccc}
\displaystyle\cos\left(\frac{s^{1-p}}{\sigma(1-p)}\right) & , & p \not= 1\\[9pt]
\cos\left(\sigma^{-1}\ln s\right)  & , & p = 1\\
\end{array}
\right. \quad \mbox{and} \quad
S_p(s) = \left\{
\begin{array}{ccc}
\displaystyle\sin\left(\frac{s^{1-p}}{\sigma(1-p)}\right) & , & p \not= 1\\[9pt]
\sin\left(\sigma^{-1}\ln s\right)  & , & p = 1\\
\end{array}
\right..
\end{equation}

If $p=\frac12$, then
\begin{equation}
\int C_{\frac12}(s)\,ds = \sigma\sqrt{s}\,S_{\frac12}(s)+\frac{\sigma^2}{2}C_{\frac12}(s)+c_1
\end{equation}
and
\begin{equation}
\int S_{\frac12}(s)\,ds = -\sigma\sqrt{s}\,C_{\frac12}(s)+\frac{\sigma^2}{2}S_{\frac12}(s)+c_2\,,
\end{equation}
where $c_1$ and $c_2$ are arbitrary constants.

Assuming for simplicity that $\mathbf{t}(s_0)=(1,0)$ and $\mathbf{n}(s_0)=(0,1)$, integration of the unit tangent, $\alpha_H=\int^s\mathbf{t}$, allows us to explicitly parametrize the hydrogen curve as
\begin{equation}
\alpha_H(s)=R_{\frac{1}{2}}(s_0)\left(
\begin{array}{cc}
\frac{\sigma^2}{2} & \sigma\sqrt{s}\\[5pt]
-\sigma\sqrt{s} & \frac{\sigma^2}{2}\\
\end{array}
\right)\left(
\begin{array}{c}
C_{\frac{1}{2}}(s)\\[5pt]
S_{\frac{1}{2}}(s)\\
\end{array}
\right)+\alpha_0\,,
\end{equation}
where $\alpha_0\in\mathbb{R}^2$ is a constant point and we have defined a ``rotation" matrix $R_p(s)$
\begin{equation}\label{eq::pRotationMatrix}
R_p(s) = \left(
\begin{array}{cc}
C_p(s) & S_p(s)\\ [5pt]
-S_p(s) & C_p(s)\\
\end{array}
\right).
\end{equation}

Note that $\Vert\alpha_H(s)-\alpha_0\Vert=\sigma\sqrt{s+\frac{\sigma^2}{4}} \sim s^{1/2}$, which shows that $\alpha_H$ spirals around a point.

\subsection{Solution of the 1D Hydrogen atom}

In the 1950s, Loudon solved the 1D Hydrogen atom on the line \cite{LoudonAmJPhys}:
\begin{equation}
    -\frac{\hbar^2}{2m}\frac{d\psi^2}{dx^2}-\frac{e^2}{\vert x\vert}\psi = E\psi,
\end{equation}
where $e$ is the electric charge of the electron and $\psi$ is complex function defined over the real line: $\psi:\mathbb{R}\to\mathbb{C}$. The difficulty of solving the 1D Hydrogen atom lies in the existence of a pole at $x=0$. The idea is to solve the equation for the regions $x<0$ and $x>0$ and then join the two solutions at $x=0$ by approaching the actual potential as the limit of a nonsingular potential $V(x)$, see, e.g., Fig. 1 of Ref. \cite{LoudonAmJPhys}.

If we write the eigenfunction along the Hydrogen curve as a function of the arc length parameter $s>0$, we have the following wave function along the curve
\begin{equation}
\psi = B{\rm e}^{-\frac{z}{2}}\,z\,L_{N}^1(z),\quad z=\frac{2s}{Na_0},
\end{equation}
where $B$ is a normalizing constant, $a_0=\hbar^2/me^2$, and $L_a^b(z)$ denotes an associated Laguerre polynomial. Note that this solution is not equal to the radial solution of the 3D Hydrogen atom:
\begin{equation}
R_{N\ell}(r) = B_{N\ell},{\rm e}^{-\frac{z}{2}}z^{\ell}L_{N}^{2\ell+1}(z),\quad z=\frac{2r}{Na_0},
\end{equation}
where $B_{N\ell}$ is a normalizing constant. However, taking into account the use of spherical coordinates to describe the radial part, one obtains the same probability density in both cases: $dP_{1D}=\vert\psi_{1D}\vert^2ds=dP_{3D}=r^2\vert\psi_{1D}\vert^2dr$, where one must take $\ell=0$ in the 3D solution to compare the solutions in both dimensions properly. As expected, this means that in the 1D solution, only $s$ orbitals make sense and, therefore, a 1D periodic table will have 2 columns only \cite{Chemistry1D,Chem1DPackage}.


\section{Polyene chains as a constrained quantum dynamics problem}
\label{sect::PolyeneCurve}

The consideration of $\pi$ electrons is essential for the stability of certain carbon compounds, such as polyenes \cite{huckel1931, penney1934}. In these compounds, $\pi$ electrons can often be approximated as particles in a box \cite{ruedenberg1953, scherr1953, platt1953}. For a particle confined in a one-dimensional box of length $L$, the solution to the Schr\"{o}dinger equation gives the wave function:

\begin{equation}
\psi_n(s) = \sqrt{\frac{2}{L}} \sin\left(\frac{n\pi s}{L}\right).
\end{equation}

The allowed energy levels, $E_n$, of the particle are quantized and given by:

\begin{equation}
E_n = \frac{n^2\pi^2\hbar^2}{2mL^2}.
\end{equation}

This model provides a good approximation for conjugated molecules with minimal alternation of bond lengths. However, for systems with significant bond-length alternation, such as long-chain polyenes, the model cannot adequately describe the finite absorption wavelength limit of the system, thus requiring adjustments \cite{autschbach2007}.

It is possible to improve the particle-in-a-box model by incorporating an effective mass into the Laplacian operator. This approach involves assessing which mass value accurately predicts the wavelengths of experimental transitions. Specifically, the calculated effective masses were $0.531m_e$, $0.457m_e$,  $0.440m_e$, and $0.384m_e$  for deca-2,4,6,8-tetraene, dodeca-2,4,6,8,10-pentaene, tetradeca-2,4,6,8,10,12-hexaene, and hexadeca-2,4,6,8,10,12,14-heptaene molecules, respectively. These values, inversely proportional to the increase in the chains' length, align with expectations from models of ballistic electrons in nanostructures and crystals, which tend towards 0.173$m_e$ \cite{santos2016}.

The $\pi$ electrons are not strongly bound to the chain, with interactions weaker than those between charges, especially noticeable in long-chain systems such as polyenes. In this work, we  propose to describe $\pi$ electrons using confinement with geometry-induced potential
\begin{equation}\label{eq::potentialsigma}
V_{\text{gip}} = - \frac{\hbar^2}{8m\sigma^2 s^2}.
\end{equation}
We aim to represent a kind of average interaction between the molecule and the $\pi$ electrons. However, this interaction would be electromagnetic, weaker than a conventional Coulomb interaction but stronger than dipole interactions or Van der Waals forces.

\subsection{Polyene curve}

Let us denote by $\alpha_P$ a plane curve whose corresponding curvature function is of the form $k=\frac{1}{\sigma s}$, i.e., Eq. \eqref{eq::PowerLawCurvature} with $p=1$. We shall refer to $\alpha_P$ as a \emph{polyene curve}.

For $p=1$, the auxiliary functions $C_p$ and $S_p$ defined in \eqref{eq::DefOfCpAndSp} have the form
\begin{equation}
\int C_{1}(s)\,ds = \frac{\sigma\,s}{1+\sigma^2}\left[S_1(s)+\sigma C_1(s)\right] + c_1
\end{equation}
and
\begin{equation}
\int S_{1}(s)\,ds = -\frac{\sigma\,s}{1+\sigma^2}\left[C_1(s)-\sigma S_1(s)\right] + c_2\,,
\end{equation}
where $c_1$ and $c_2$ are arbitrary real constants.

Assuming for simplicity that $\mathbf{t}(s_0)=(1,0)$ and $\mathbf{n}(s_0)=(0,1)$, integration of the unit tangent, $\alpha_P=\int^s\mathbf{t}$, allows us to explicitly parametrize the polyene curve as:
\begin{equation}\label{eq::Alpha_P}
\alpha_P(s)=\frac{\sigma s}{1+\sigma^2}\,R_{1}(s_0)\left(
\begin{array}{cr}
\sigma & 1 \\
1 & -\sigma\\
\end{array}
\right)\left(
\begin{array}{c}
C_{1}(s)\\
S_{1}(s)\\
\end{array}
\right)+\alpha_0\,,
\end{equation}
where $\alpha_0\in\mathbb{R}^2$ is constant and $R_1(s)$ is defined as in Eq. \eqref{eq::pRotationMatrix}.

Note that $\Vert\alpha_P(s)-\alpha_0\Vert=\frac{\sigma}{\sqrt{1+\sigma^2}} s$, which shows that $\alpha_P$ spirals around a point, as happens for the hydrogen curve. Note that polyene curves approach its initial point $\alpha_0$ faster than the hydrogen curve: $\frac{\Vert\alpha_P(s)-\alpha_0\Vert}{\Vert\alpha_H(s)-\alpha_0\Vert} \overset{s\to0}{\longrightarrow}0$. In addition, polyene curves rotate around its initial point $\alpha_0$ more than the hydrogen curve, as a comparison between $\{C_{1}(s),S_{1}(s)\}$ and $\{C_{\frac{1}{2}}(s),S_{\frac{1}{2}}(s)\}$ indicates.


\subsection{Spectrum of $\pi$ electrons in polyene chains}
\label{sect::PolyeneSpectrum}

\begin{figure}[t]
\centering
\includegraphics[width=0.8\textwidth]{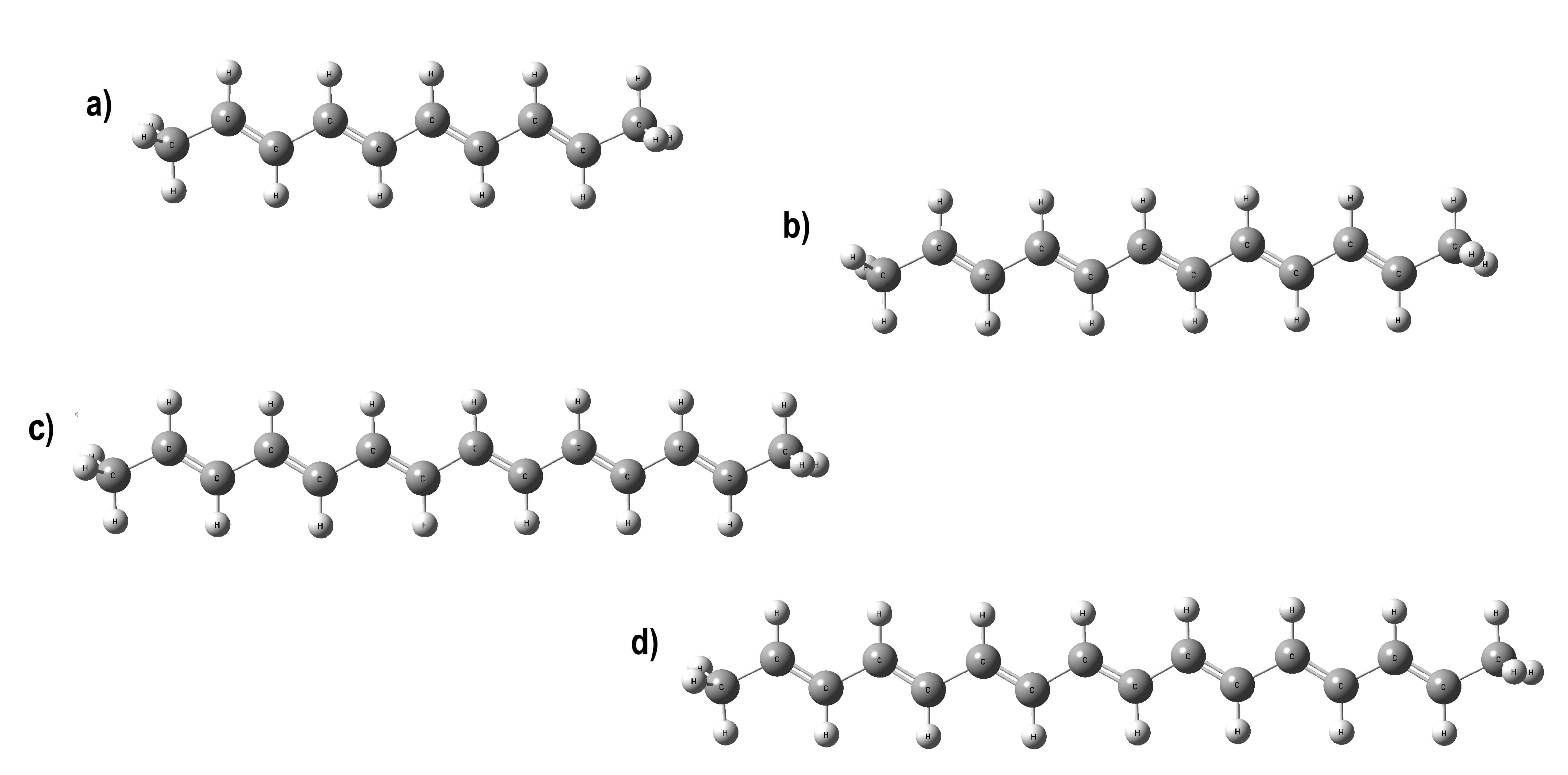}
\caption{\centering Polyenes. a) deca-2,4,6,8-tetraene; b) dodeca-2,4,6,8,10-pentaene; c) tetradeca-2,4,6,8,10,12-hexaene; d) hexadeca-2,4,6,8,10,12,14-heptaene.}
\label{fig:polyenes}
\end{figure}

In this subsection, we provide an alternative description of $\pi$ electrons confined in a 1D curved box, as defined by the polyene curves discussed in the previous subsection. Bessel functions exhibit more complex behaviour than usual trigonometric functions and introduce new factors such as zero modes \cite{kobayashi2002} and the dependence of wave amplitude on the energy level. We apply Eq. \eqref{eq::Hgip1D} to describe the $\pi$ electrons in the polyenes deca-2,4,6,8-tetraene, dodeca-2,4,6,8,10-pentaene, tetradeca-2,4,6,8,10,12-hexaene, and hexadeca-2,4,6,8,10,12,14-heptaene \citep{Christensen2008} (Fig. \ref{fig:polyenes}).

The polyene curve parametrization is given by:
\begin{equation}
\alpha_P(s) = \frac{\sigma s}{1+\sigma^2}\Big(\cos\left(\frac{\ln s}{\sigma}\right) + \sigma \sin\left(\frac{\ln s}{\sigma}\right),\sin\left( \frac{\ln s}{\sigma}\right) - \sigma \cos \left(\frac{\ln s}{\sigma}\right)\Big),
\end{equation}
where we have set $s_0=1$ and $\alpha_0=(0,0)$ in Eq. \eqref{eq::Alpha_P}.

For the geometry-induced potential of polyenes curves, $V_{gip}=-\frac{\hbar^2}{8m\sigma^2 s^2}$, the equation \eqref{eq::Hgip1D} becomes
\begin{equation}
    -\frac{d^2\psi}{ds^2} = \left(\epsilon + \frac{1}{4\sigma^2s^2}\right)\psi, \quad \epsilon = \frac{2mE}{\hbar^2},
\end{equation}
whose general solution is expressed as a linear combination of Bessel functions $J_{\omega}(s)$ and $Y_{\omega}(s)$ of the first and second types, respectively:
\begin{equation}\label{eq::psi}
\psi_n (s) = c_1 \sqrt{s}\,J_{\omega}(\sqrt{\epsilon}\,s) + c_2 \sqrt{s}\,Y_{\omega}(\sqrt{\epsilon}\,s), \quad \omega = \frac{1}{2}\sqrt{\left\vert 1-\frac{1}{\sigma^2}\right\vert}\,.
\end{equation}

We want a solution on the interval $[0,L]$ and, therefore, must impose the condition $c_2=0$. (Bessel functions of the second type diverge at the origin.) Applying the homogeneous boundary conditions in Eq. \eqref{eq::psi}, the solutions are subject to the relationship:

\begin{equation}
L = \frac{j_{\omega, n}}{\sqrt{\epsilon}} \quad \Rightarrow \quad E_n = \frac{\hbar^2}{2mL^2}\,j_{\omega, n}^2\,,
\end{equation}

where $j_{\omega, n}$ denotes the $n$-th zero of $J_{\omega}$.

To determine the value of $c_1$, we can use the normalization condition

\begin{equation}
1 = \int_{0}^{L} |\psi_n (s)|^2 \,ds = c_1^2\int_{0}^{L} |\sqrt{s}\,J_{\omega}(\sqrt{\epsilon}\,s)|^2 \,ds,
\end{equation}
and obtain
\begin{equation}
 c_1 = \frac{\sqrt{2}}{L\sqrt{J_{\omega}(\sqrt{\epsilon}\,s) - J_{\omega-1}(\sqrt{\epsilon}\,s)J_{\omega+1}(\sqrt{\epsilon}\,s)]}}.
\end{equation}

Using that $\epsilon = \frac{j_{\omega, n}}{L}$ and $J_{\omega}(j_{\omega, n}) = 0$, we have

\begin{equation}
c_1 = \frac{\sqrt{2}}{L\sqrt{- J_{\omega-1}(j_{\omega, n})J_{\omega+1}(j_{\omega, n})}}.
\end{equation}

Thus, the complete basis of wave functions is
\begin{equation}
\psi_n (s) = \frac{\sqrt{2}}{L\sqrt{-J_{\omega-1}(j_{\omega, n})J_{\omega+1}(j_{\omega, n})}}\,\sqrt{s}\,J_{\omega}\left(\frac{j_{\omega, n}}{L}s\right).
\end{equation}

\begin{table}[t]
    \centering
    \caption{\centering Values of $\sigma$ and $\omega$ for distinct polyene chains.}
    \label{tab::k1SigmaOmega}
    \small
    \begin{tabular}{ccccccc}
        \toprule
        \textbf{System} & \textbf{$\sigma$} & \textbf{$\omega$} &   \\
        \midrule
        deca-2,4,6,8-tetraene & $\sqrt{0.004}$ & 7.88987  \\
        dodeca-2,4,6,8,10-pentaene & $\sqrt{0.0014}$ & 13.35370  \\
        tetradeca-2,4,6,8,10,12-hexaene & $\sqrt{0.0009}$ & 16.65920  \\
        hexadeca-2,4,6,8,10,12,14-heptaene & $\sqrt{0.00045}$ & 23.56490  \\
        \bottomrule
    \end{tabular}
\end{table}

For each polyene, specific values of $\sigma$ and $\omega$ were obtained, representing system structure changes. Table \ref{tab::k1SigmaOmega} shows these values for four distinct systems. The value of $\sigma$ was determined using the relationship between the wavelength of transition $\lambda$ and the energy change $\Delta E$: $\lambda = \frac{hc}{\Delta E}$, where $c$ is the velocity of light in vacuum and $h$ the Planck constant. Comparison with experimental values of $\lambda$ \cite{Christensen2008} allows us to estimate the corresponding values of $\sigma$. Note that the appropriate energy levels used to compute $\Delta E$ depend on how many $\pi$ electrons the polyene chains have, namely, 8 for the deca-2,4,6,8-tetraene molecule ($n = 4$), 10 for the dodeca-2,4,6,8,10-pentaene molecule ($n = 5$), 12 for the tetradeca-2,4,6,8,10,12-hexaene molecule ($n = 6$), and 14 for the hexadeca-2,4,6,8,10,12,14-heptaene molecule ($n = 7$).

\begin{figure}[t]%
\centering
\begin{subfigure}{0.4\textwidth}
  \centering
  \includegraphics[width=\textwidth]{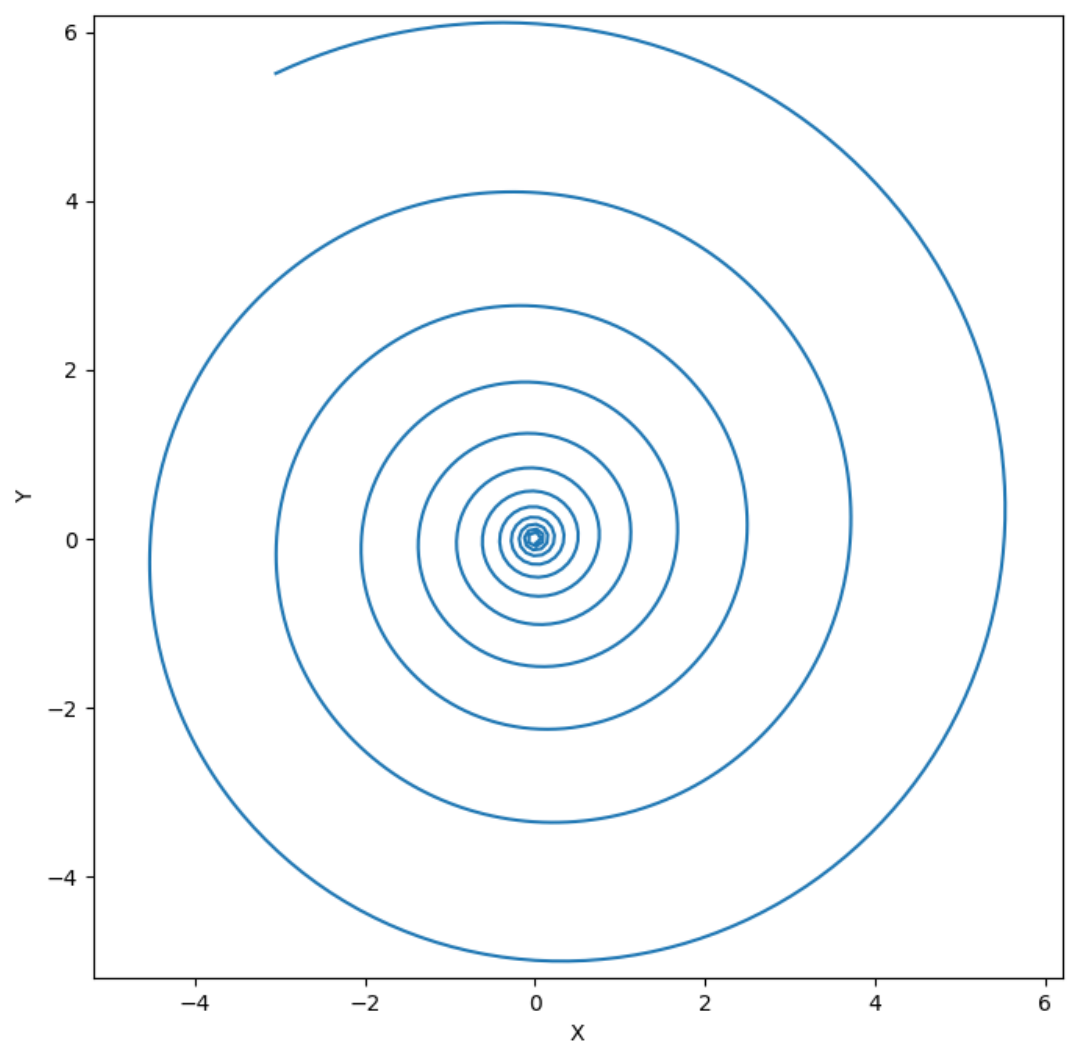}
  \caption{$\sigma=\sqrt{0.004}$}
  \label{fig:sigma004}
\end{subfigure}
\begin{subfigure}{0.4\textwidth}
  \centering
  \includegraphics[width=\textwidth]{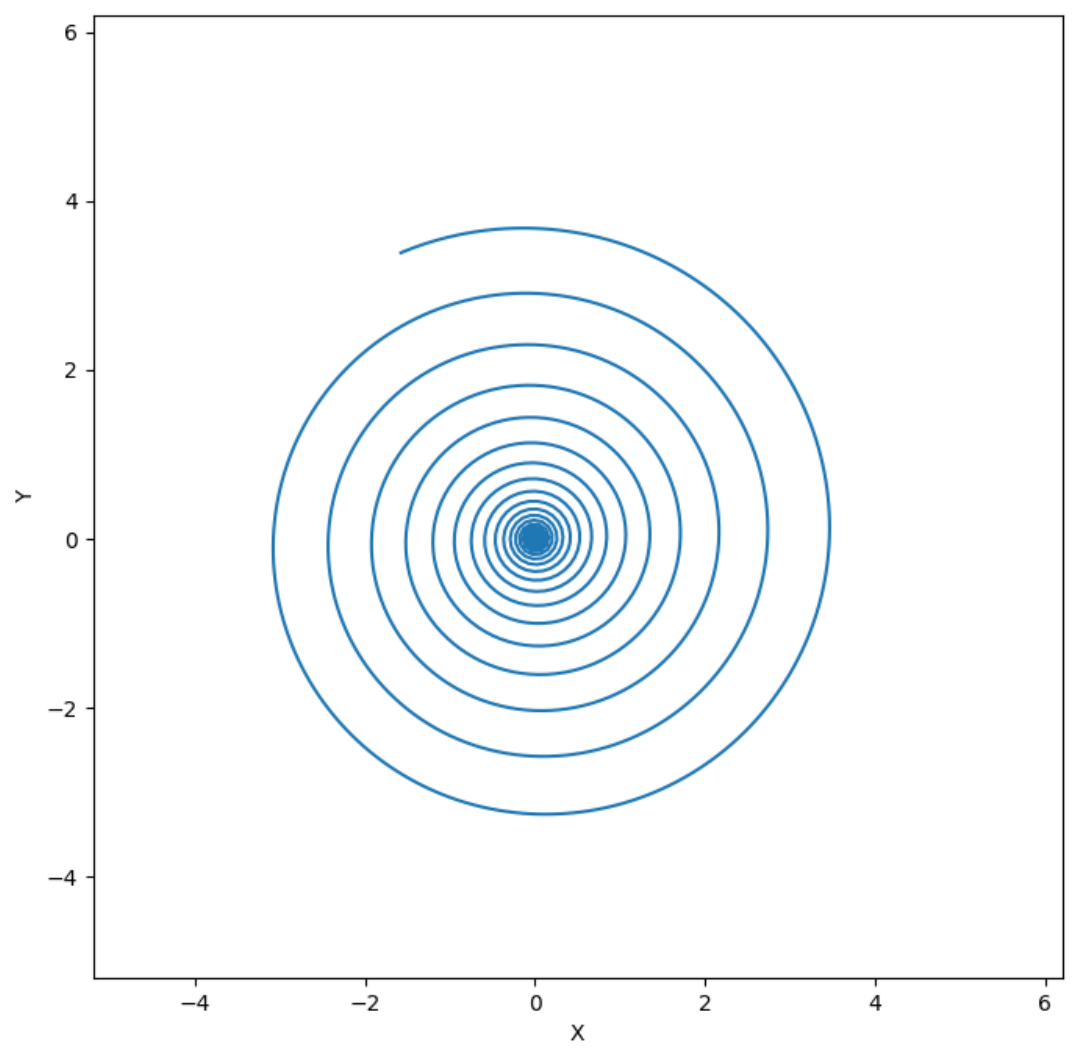}
  \caption{$\sigma=\sqrt{0.0014}$}
  \label{fig:sigma0014}
\end{subfigure}
\begin{subfigure}{0.4\textwidth}
  \centering
  \includegraphics[width=\textwidth]{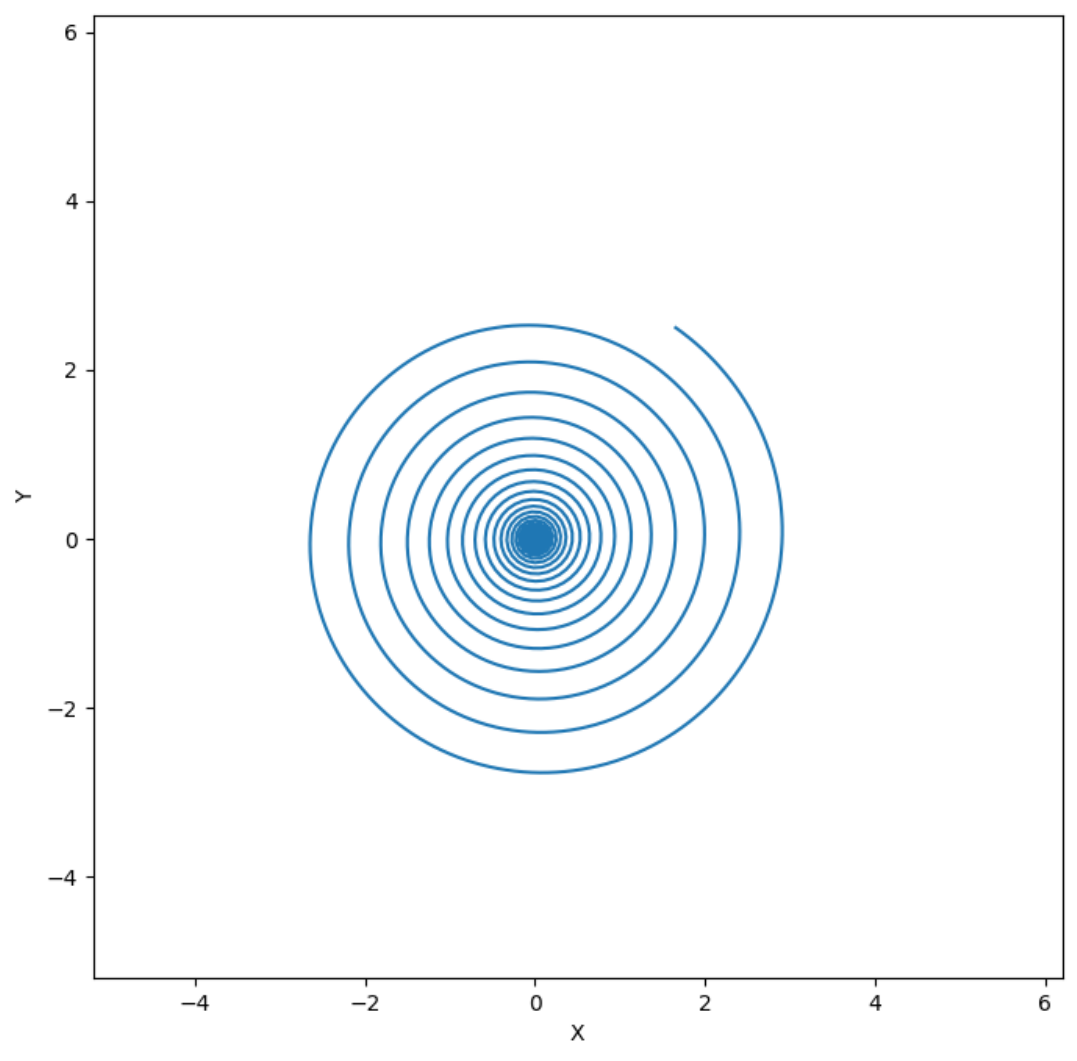}
  \caption{$\sigma=\sqrt{0.0009}$}
  \label{fig:sigma0009}
\end{subfigure}
\begin{subfigure}{0.4\textwidth}
  \centering
  \includegraphics[width=\textwidth]{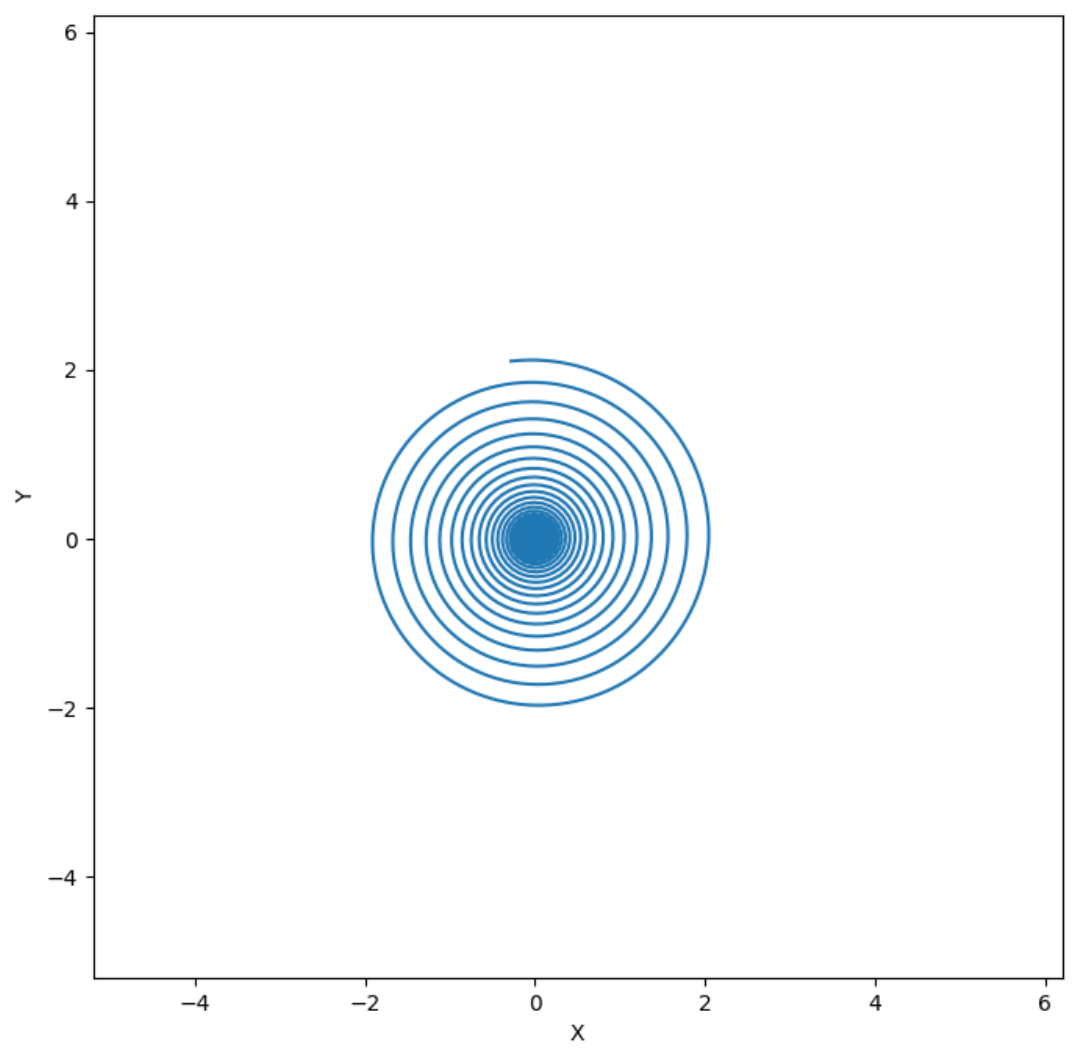}
  \caption{$\sigma=\sqrt{0.00045}$}
  \label{fig:sigma00045}
\end{subfigure}
\caption{Plots of polyene curves, i.e., curves with curvature function $k(s)=\frac{1}{\sigma s}$. a) deca-2,4,6,8-tetraene; b) dodeca-2,4,6,8,10-pentaene; c) tetradeca-2,4,6,8,10,12-hexaene; d) hexadeca-2,4,6,8,10,12,14-heptaene.}
\label{fig:spirals}
\end{figure}

\begin{figure}[t]%
\centering
\includegraphics[width=0.8\textwidth]{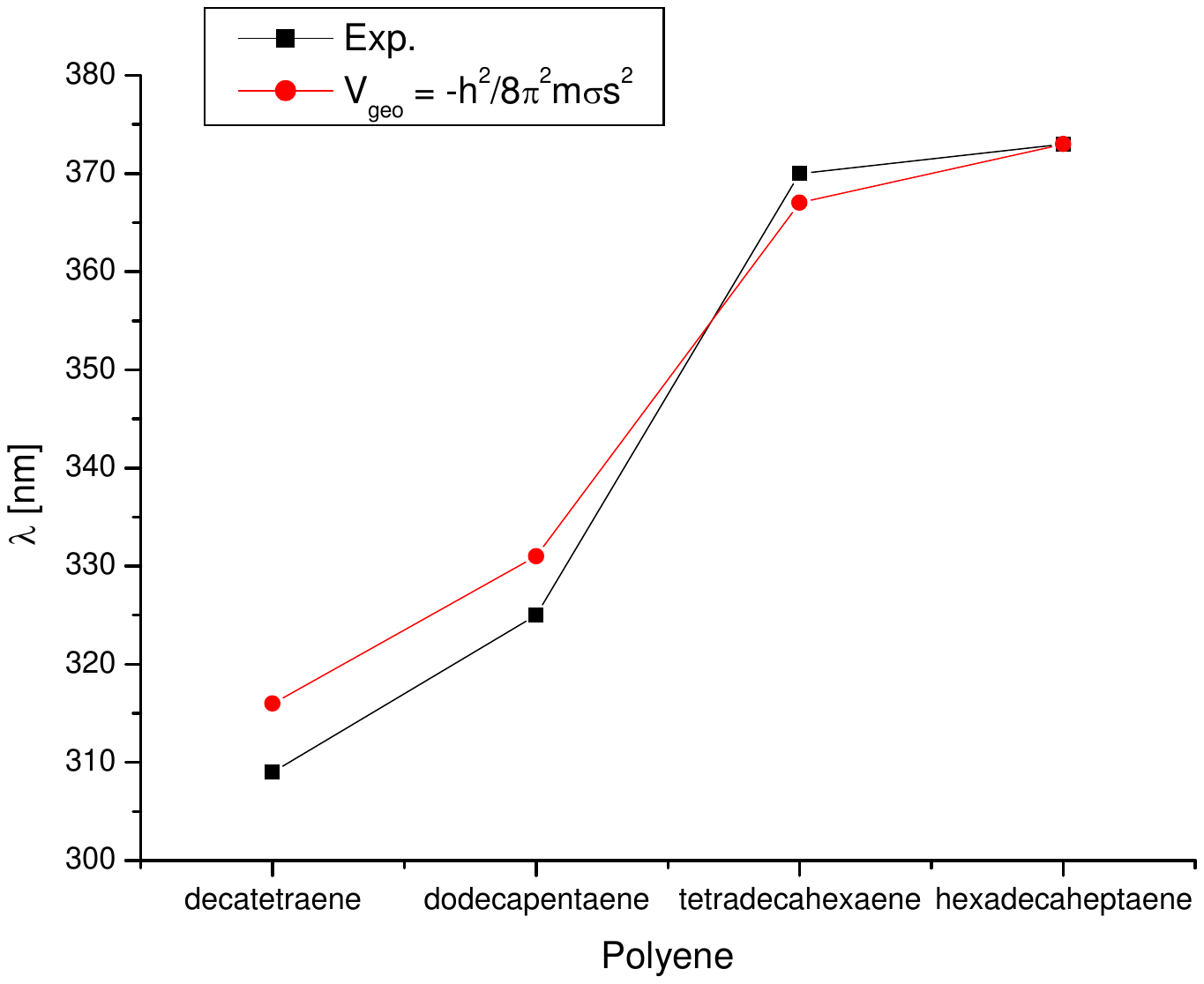}
\caption{\centering Adjustment of the parameter $\sigma$ for each polyene curve.}
\label{fig:calcvsexp}
\end{figure}

The polyene curves, i.e., plane curves with curvature $k(s) = \frac{1}{\sigma s}$, are depicted in Fig. \ref{fig:spirals}. These spiral curves have a curvature function that resembles that used to simulate the hydrogen atom (Section \ref{sect::1dHatom}). Power-law potentials $V \propto s^{-p}$ also describe interactions such as charge-charge, dipole-dipole, and also Van der Waals, a very important class of potentials in chemistry. The relationship between $\sigma$, the number of $\pi$ electrons, and the spiral geometry seems to be established, as it is evident that the curve's pitch, i.e., the distance between two points after a revolution of the spiral, tends to decrease with the reduction of $\sigma$ and the increase in the number of $\pi$ electrons. In other words, $\sigma$ approaches zero with an increase in the number of $\pi$ electrons. This is equivalent to an increase in the geometry-induced potential, justifying the decrease in energy.

In the case of the Hydrogen atom, the direction toward the spiral's centre indicates the nucleus's attraction. However, for $\pi$ electrons, this association of potential to geometry is not as direct. Comparing our results with the experimental absorption values calculated using $V_{gip}$ of polyenes, we observe good agreement (Fig. \ref{fig:calcvsexp}). The errors obtained are 2.30\% for deca-2,4,6,8-tetraene, 1.85\% for dodeca-2,4,6,8,10-pentaene, 0.81\% for tetradeca-2,4,6,8,10,12-hexaene, and 0.00\% for hexadeca-2,4,6,8,10,12,14-heptaene.

The figure \ref{fig:calcvsexp} shows that the $\pi$-$\pi^*$ transitions are satisfactorily described with the geometry-induced potential. It is worth noting that accurate results for these transitions can also be obtained with a particle-in-a-box model using an appropriate effective mass.


\section{Conclusion}
\label{sect::Conclusion}

In this work, we employed a method to obtain the $\pi$ electron spectrum using the Jensen-Koppe-Da Costa's confining potential formalism. Treating $\pi$ electrons as trapped in a 1D curved spiral box leads to a Schrödinger equation subjected a geometry-induced potential depending on the geometry of the corresponding spiral curve. The analytically obtained spectrum allowed us to describe the $\pi$-$\pi^*$ transitions of the polyenes chains deca-2,4,6,8-tetraene, dodeca-2,4,6,8,10-pentaene, tetradeca-2,4,6,8,10,12-hexaene, and hexadeca-2,4,6,8,10,12,14-heptaene. The solutions are given by Bessel functions, which provide descriptions beyond the particle-in-a-box model. Studying different systems with $\pi$ resonances should demonstrate new uses of geometry-induced potential.

\section*{ACKNOWLEDGMENT}\label{sec5}

We acknowledge the Coordenação de Aperfeiçoamento de Pessoal de Nível Superior (CAPES) for their financial support of this work.


\begin{thebibliography}{24}
\ifx \bisbn   \undefined \def \bisbn  #1{ISBN #1}\fi
\ifx \binits  \undefined \def \binits#1{#1}\fi
\ifx \bauthor  \undefined \def \bauthor#1{#1}\fi
\ifx \batitle  \undefined \def \batitle#1{#1}\fi
\ifx \bjtitle  \undefined \def \bjtitle#1{#1}\fi
\ifx \bvolume  \undefined \def \bvolume#1{\textbf{#1}}\fi
\ifx \byear  \undefined \def \byear#1{#1}\fi
\ifx \bissue  \undefined \def \bissue#1{#1}\fi
\ifx \bfpage  \undefined \def \bfpage#1{#1}\fi
\ifx \blpage  \undefined \def \blpage #1{#1}\fi
\ifx \burl  \undefined \def \burl#1{\textsf{#1}}\fi
\ifx \doiurl  \undefined \def \doiurl#1{\url{https://doi.org/#1}}\fi
\ifx \betal  \undefined \def \betal{\textit{et al.}}\fi
\ifx \binstitute  \undefined \def \binstitute#1{#1}\fi
\ifx \binstitutionaled  \undefined \def \binstitutionaled#1{#1}\fi
\ifx \bctitle  \undefined \def \bctitle#1{#1}\fi
\ifx \beditor  \undefined \def \beditor#1{#1}\fi
\ifx \bpublisher  \undefined \def \bpublisher#1{#1}\fi
\ifx \bbtitle  \undefined \def \bbtitle#1{#1}\fi
\ifx \bedition  \undefined \def \bedition#1{#1}\fi
\ifx \bseriesno  \undefined \def \bseriesno#1{#1}\fi
\ifx \blocation  \undefined \def \blocation#1{#1}\fi
\ifx \bsertitle  \undefined \def \bsertitle#1{#1}\fi
\ifx \bsnm \undefined \def \bsnm#1{#1}\fi
\ifx \bsuffix \undefined \def \bsuffix#1{#1}\fi
\ifx \bparticle \undefined \def \bparticle#1{#1}\fi
\ifx \barticle \undefined \def \barticle#1{#1}\fi
\bibcommenthead
\ifx \bconfdate \undefined \def \bconfdate #1{#1}\fi
\ifx \botherref \undefined \def \botherref #1{#1}\fi
\ifx \url \undefined \def \url#1{\textbf{#1}}\fi
\ifx \bchapter \undefined \def \bchapter#1{#1}\fi
\ifx \bbook \undefined \def \bbook#1{#1}\fi
\ifx \bcomment \undefined \def \bcomment#1{#1}\fi
\ifx \oauthor \undefined \def \oauthor#1{#1}\fi
\ifx \citeauthoryear \undefined \def \citeauthoryear#1{#1}\fi
\ifx \endbibitem  \undefined \def \endbibitem {}\fi
\ifx \bconflocation  \undefined \def \bconflocation#1{#1}\fi
\ifx \arxivurl  \undefined \def \arxivurl#1{\textsf{#1}}\fi
\csname PreBibitemsHook\endcsname
\bibitem[\protect\citeauthoryear{Mannix et~al.}{2017}]{Chem2D2017}
\begin{barticle}
\bauthor{\bsnm{Mannix}, \binits{A.J.}},
\bauthor{\bsnm{Kiraly}, \binits{B.}},
\bauthor{\bsnm{Hersam}, \binits{M.C.}},
\bauthor{\bsnm{Guisinger}, \binits{N.P.}}:
\batitle{Synthesis and chemistry of elemental {2D} materials}.
\bjtitle{Nature Rev. Chem.}
\bvolume{1}(\bissue{2}),
\bfpage{0014}
(\byear{2017})
\doiurl{10.1038/s41570-016-0014}
\end{barticle}
\endbibitem
\bibitem[\protect\citeauthoryear{De~Witt}{1957}]{dW57}
\begin{barticle}
\bauthor{\bsnm{De~Witt}, \binits{B.S.}}:
\batitle{Dynamical theory in curved spaces. {I}. a review of the classical and quantum action principles}.
\bjtitle{Rev. Mod. Phys.}
\bvolume{29},
\bfpage{377}
(\byear{1957})
\doiurl{10.1103/RevModPhys.29.377}
\end{barticle}
\endbibitem
\bibitem[\protect\citeauthoryear{Jensen and Koppe}{1971}]{JensenKoppe1971}
\begin{barticle}
\bauthor{\bsnm{Jensen}, \binits{H.}},
\bauthor{\bsnm{Koppe}, \binits{H.}}:
\batitle{Quantum mechanics with constraints}.
\bjtitle{Annals of Physics}
\bvolume{63},
\bfpage{586}--\blpage{591}
(\byear{1971})
\doiurl{10.1016/0003-4916(71)90031-5}
\end{barticle}
\endbibitem
\bibitem[\protect\citeauthoryear{da~Costa}{1981}]{DaCosta1981}
\begin{barticle}
\bauthor{\bsnm{da Costa}, \binits{R.C.T.}}:
\batitle{Quantum mechanics of a constrained particle}.
\bjtitle{Phys. Rev. A}
\bvolume{23},
\bfpage{1982}
(\byear{1981})
\doiurl{10.1103/PhysRevA.23.1982}
\end{barticle}
\endbibitem
\bibitem[\protect\citeauthoryear{{del Campo} et~al.}{2014}]{DelCampo2014}
\begin{barticle}
\bauthor{\bsnm{{del Campo}}, \binits{A.}},
\bauthor{\bsnm{Boshier}, \binits{M.G.}},
\bauthor{\bsnm{Saxena}, \binits{A.}}:
\batitle{Bent waveguides for matter-waves: supersymmetric potentials and reflectionless geometries}.
\bjtitle{Sci. Rep.}
\bvolume{4},
\bfpage{5274}
(\byear{2014})
\doiurl{10.1038/srep05274}
\end{barticle}
\endbibitem
\bibitem[\protect\citeauthoryear{{da Silva} et~al.}{2017}]{DaSilva2017}
\begin{barticle}
\bauthor{\bsnm{{da Silva}}, \binits{L.C.B.}},
\bauthor{\bsnm{Bastos}, \binits{C.C.}},
\bauthor{\bsnm{Ribeiro}, \binits{F.G.}}:
\batitle{Quantum mechanics of a constrained particle and the problem of prescribed geometry-induced potential}.
\bjtitle{Annals of Physics}
\bvolume{379},
\bfpage{13}--\blpage{33}
(\byear{2017})
\doiurl{10.1016/j.aop.2017.02.012}
\end{barticle}
\endbibitem
\bibitem[\protect\citeauthoryear{Lima et~al.}{2021}]{Lima2021}
\begin{barticle}
\bauthor{\bsnm{Lima}, \binits{J.D.M.}},
\bauthor{\bsnm{Gomes}, \binits{E.}},
\bauthor{\bsnm{{da Silva Filho}}, \binits{F.F.}},
\bauthor{\bsnm{Moraes}, \binits{F.}},
\bauthor{\bsnm{Teixeira}, \binits{R.}}:
\batitle{Geometric effects on the electronic structure of curved nanotubes and curved graphene: the case of the helix, catenary, helicoid, and catenoid}.
\bjtitle{Eur. Phys. J. Plus}
\bvolume{136},
\bfpage{551}
(\byear{2021})
\doiurl{10.1140/epjp/s13360-021-01533-6}
\end{barticle}
\endbibitem
\bibitem[\protect\citeauthoryear{Onoe et~al.}{2012}]{Onoe12GeomEffectsPeanutTube}
\begin{barticle}
\bauthor{\bsnm{Onoe}, \binits{J.}},
\bauthor{\bsnm{Ito}, \binits{T.}},
\bauthor{\bsnm{Shima}, \binits{H.}},
\bauthor{\bsnm{Yoshioka}, \binits{H.}},
\bauthor{\bsnm{Kimura}, \binits{S.-I.}}:
\batitle{Observation of Riemannian geometric effects on electronic states}.
\bjtitle{Europhys. Lett.}
\bvolume{98}(\bissue{2}),
\bfpage{27001}
(\byear{2012})
\doiurl{10.1209/0295-5075/98/2700}
\end{barticle}
\endbibitem
\bibitem[\protect\citeauthoryear{Bastos et~al.}{2016}]{Bastos2016}
\begin{barticle}
\bauthor{\bsnm{Bastos}, \binits{C.C.}},
\bauthor{\bsnm{Pavão}, \binits{A.C.}},
\bauthor{\bsnm{Leandro}, \binits{E.S.G.}}:
\batitle{On the quantum mechanics of a particle constrained to generalized cylinders with application to {M}öbius strips}.
\bjtitle{J. Math. Chem.}
\bvolume{54},
\bfpage{1822}--\blpage{1834}
(\byear{2016})
\doiurl{10.1007/s10910-016-0652-5}
\end{barticle}
\endbibitem
\bibitem[\protect\citeauthoryear{Oshikiri et~al.}{1996}]{Oshikiri1996}
\begin{barticle}
\bauthor{\bsnm{Oshikiri}, \binits{M.}},
\bauthor{\bsnm{Takehana}, \binits{K.}},
\bauthor{\bsnm{Asano}, \binits{T.}},
\bauthor{\bsnm{Kido}, \binits{G.}}:
\batitle{Far-infrared cyclotron resonance of wide-gap semiconductors using pulsed high magnetic fields}.
\bjtitle{Physica B: Condensed Matter}
\bvolume{216}(\bissue{3-4}),
\bfpage{351}--\blpage{353}
(\byear{1996})
\doiurl{10.1016/0921-4526(95)00515-3}
\end{barticle}
\endbibitem
\bibitem[\protect\citeauthoryear{Ruedenberg and Scherr}{1953}]{ruedenberg1953}
\begin{barticle}
\bauthor{\bsnm{Ruedenberg}, \binits{K.}},
\bauthor{\bsnm{Scherr}, \binits{C.W.}}:
\batitle{Free‐electron network model for conjugated systems. {I}. {T}heory}.
\bjtitle{J. Chem. Phys.}
\bvolume{21}(\bissue{9}),
\bfpage{1565}--\blpage{1581}
(\byear{1953})
\doiurl{10.1063/1.1699299}
\end{barticle}
\endbibitem
\bibitem[\protect\citeauthoryear{Scherr}{1953}]{scherr1953}
\begin{barticle}
\bauthor{\bsnm{Scherr}, \binits{C.W.}}:
\batitle{Free‐electron network model for conjugated systems. {II}. {N}umerical calculations}.
\bjtitle{J. Chem. Phys.}
\bvolume{21}(\bissue{9}),
\bfpage{1582}--\blpage{1596}
(\byear{1953})
\doiurl{10.1063/1.1699300}
\end{barticle}
\endbibitem
\bibitem[\protect\citeauthoryear{Platt}{1953}]{platt1953}
\begin{barticle}
\bauthor{\bsnm{Platt}, \binits{J.R.}}:
\batitle{Free‐electron network model for conjugated systems. {III}. {A} demonstration model showing bond order and ``free valence'' in conjugated hydrocarbons}.
\bjtitle{J. Chem. Phys.}
\bvolume{21}(\bissue{9}),
\bfpage{1597}--\blpage{1600}
(\byear{1953})
\doiurl{10.1063/1.1699301}
\end{barticle}
\endbibitem
\bibitem[\protect\citeauthoryear{Bastos et~al.}{2012}]{Bastos2012}
\begin{barticle}
\bauthor{\bsnm{Bastos}, \binits{C.C.}},
\bauthor{\bsnm{Paiva}, \binits{G.S.}},
\bauthor{\bsnm{Leandro}, \binits{E.S.G.}},
\bauthor{\bsnm{Pav\~{a}o}, \binits{A.C.}}:
\batitle{An extension of the particle in a one dimensional box model}.
\bjtitle{Phys. Education}
\bvolume{28},
\bfpage{1}
(\byear{2012})
\end{barticle}
\endbibitem
\bibitem[\protect\citeauthoryear{do~Carmo}{1976}]{DoCarmo1976}
\begin{bbook}
\bauthor{\bsnm{Carmo}, \binits{M.P.}}:
\bbtitle{Differential Geometry of Curves and Surfaces}.
\bpublisher{Prentice Hall},
\blocation{New Jersey}
(\byear{1976})
\end{bbook}
\endbibitem
\bibitem[\protect\citeauthoryear{Loudon}{1959}]{LoudonAmJPhys}
\begin{barticle}
\bauthor{\bsnm{Loudon}, \binits{R.}}:
\batitle{One-dimensional hydrogen atom}.
\bjtitle{Am. J. Phys.}
\bvolume{27},
\bfpage{649}
(\byear{1959})
\doiurl{10.1119/1.1934950}
\end{barticle}
\endbibitem
\bibitem[\protect\citeauthoryear{Loos et~al.}{2015}]{Chemistry1D}
\begin{barticle}
\bauthor{\bsnm{Loos}, \binits{P.-F.}},
\bauthor{\bsnm{Ball}, \binits{C.J.}},
\bauthor{\bsnm{Gill}, \binits{P.M.W.}}:
\batitle{Chemistry in one dimension}.
\bjtitle{Phys. Chem. Chem. Phys.}
\bvolume{17},
\bfpage{3196}
(\byear{2015})
\doiurl{10.1039/C4CP03571B}
\end{barticle}
\endbibitem
\bibitem[\protect\citeauthoryear{Ball and Gill}{2015}]{Chem1DPackage}
\begin{barticle}
\bauthor{\bsnm{Ball}, \binits{C.J.}},
\bauthor{\bsnm{Gill}, \binits{P.M.W.}}:
\batitle{{Chem1D}: a software package for electronic structure calculations on one-dimensional systems}.
\bjtitle{Mol. Phys.}
\bvolume{113},
\bfpage{1843}
(\byear{2015})
\doiurl{10.1080/00268976.2015.1017018}
\end{barticle}
\endbibitem
\bibitem[\protect\citeauthoryear{Hückel}{1931}]{huckel1931}
\begin{barticle}
\bauthor{\bsnm{Hückel}, \binits{E.}}:
\batitle{Quantentheoretische Beiträge zum Benzolproblem}.
\bjtitle{Zeitschrift für Physik}
\bvolume{70},
\bfpage{204}--\blpage{286}
(\byear{1931})
\doiurl{10.1007/BF01339530}
\end{barticle}
\endbibitem
\bibitem[\protect\citeauthoryear{Penney}{1934}]{penney1934}
\begin{barticle}
\bauthor{\bsnm{Penney}, \binits{W.G.}}:
\batitle{The Theory of the Stability of the Benzene Ring and Related Compounds}.
\bjtitle{Proceedings of the Royal Society A: Mathematical, Physical and Engineering Sciences}
\bvolume{146}(\bissue{856}),
\bfpage{223}--\blpage{238}
(\byear{1934})
\doiurl{10.1098/rspa.1934.0151}
\end{barticle}
\endbibitem
\bibitem[\protect\citeauthoryear{Autschbach}{2007}]{autschbach2007}
\begin{barticle}
\bauthor{\bsnm{Autschbach}, \binits{J.}}:
\batitle{Why the Particle-in-a-Box Model Works Well for Cyanine Dyes but Not for Conjugated Polyenes}.
\bjtitle{Journal of Chemical Education}
\bvolume{84}(\bissue{11}),
\bfpage{1840}
(\byear{2007})
\doiurl{10.1021/ed084p1840}
\end{barticle}
\endbibitem
\bibitem[\protect\citeauthoryear{Santos et~al.}{2016}]{santos2016}
\begin{barticle}
\bauthor{\bsnm{Santos}, \binits{F.}},
\bauthor{\bsnm{Fumeron}, \binits{S.}},
\bauthor{\bsnm{Berche}, \binits{B.}},
\bauthor{\bsnm{Moraes}, \binits{F.}}:
\batitle{Geometric effects in the electronic transport of deformed nanotubes}.
\bjtitle{Nanotechnology}
\bvolume{27},
\bfpage{135302}
(\byear{2016})
\doiurl{10.1088/0957-4484/27/13/135302}
\end{barticle}
\endbibitem
\bibitem[\protect\citeauthoryear{Kobayashi and Shimbori}{2002}]{kobayashi2002}
\begin{botherref}
\bauthor{\bsnm{Kobayashi}, \binits{T.}},
\bauthor{\bsnm{Shimbori}, \binits{T.}}:
\batitle{Zero-energy solutions and vortices in {S}chr{\"o}dinger equations}.
\bjtitle{Phys. Rev. A}
\bvolume{65},
\bfpage{042108}
(\byear{2002})
\doiurl{10.1103/physreva.65.042108}
\end{botherref}
\endbibitem
\bibitem[\protect\citeauthoryear{Christensen et~al.}{2008}]{Christensen2008}
\begin{barticle}
\bauthor{\bsnm{Christensen}, \binits{R.L.}},
\bauthor{\bsnm{Galinato}, \binits{M.G.I.}},
\bauthor{\bsnm{Chu}, \binits{E.F.}},
\bauthor{\bsnm{Howard}, \binits{J.N.}},
\bauthor{\bsnm{Broene}, \binits{R.D.}},
\bauthor{\bsnm{Frank}, \binits{H.A.}}:
\batitle{Energies of low-lying excited states of linear polyenes}.
\bjtitle{J. Phys. Chem. A}
\bvolume{112},
\bfpage{12629}--\blpage{12636}
(\byear{2008})
\doiurl{10.1021/jp8060202}
\end{barticle}
\endbibitem

\end{thebibliography}
\end{document}